# Berry curvature generation detected by Nernst responses in ferroelectric Weyl semimetal

Cheng-Long Zhang[*,1] Tian Liang[*,1] M. S. Bahramy[*,2] Naoki Ogawa,[1,3,4] Vilmos Kocsis,[1] Kentaro Ueda,[3] Yoshio Kaneko,[1] Markus Kriener,[1] Yoshinori Tokura[1,3,5]

[1]*RIKEN Center for Emergent Matter Science, Wako 351-0198, Japan*

[2]*Department of Physics and Astronomy, The University of Manchester, Oxford Road, Manchester M13 9PL, United Kingdom*

[3]*Department of Applied Physics, University of Tokyo, Tokyo 113-8656, Japan*

[4]*PRESTO, Japan Science and Technology Agency (JST), 332-0012, Kawaguchi, Japan*

[5]*Tokyo College, University of Tokyo, Tokyo 113-8656, Japan*

*\* These authors contributed equally.*

*Corresponding authors (emails): chenglong.zhang@riken.jp; tian.liang@riken.jp*

The quest for non-magnetic Weyl semimetals with high tunability of phase has remained a demanding challenge. As the symmetry-breaking control parameter, the ferroelectric order can be steered to turn on/off the Weyl semimetals phase, adjust the band structures around the Fermi level, and enlarge/shrink the momentum separation of Weyl nodes which generate the Berry curvature as the emergent magnetic field. Here, we report the realization of a ferroelectric non-magnetic Weyl semimetal based on indium-doped $Pb_{1-x}Sn_xTe$ alloy where the underlying inversion symmetry as well as mirror symmetry is broken with the strength of ferroelectricity adjustable via tuning indium doping level and Sn/Pb ratio. The transverse thermoelectric effect, i.e., Nernst effect both for out-of-plane and in-plane magnetic-field geometry, is exploited as a Berry-curvature-sensitive experimental probe to manifest the generation of Berry curvature via the re-distribution of Weyl nodes under magnetic fields. The results demonstrate a clean non-magnetic Weyl semimetal coupled with highly tunable ferroelectric order, providing an ideal platform for manipulating the Weyl fermions in non-magnetic system.

Weyl nodes, which serve as the sources of Berry curvature, can appear in a solid system when it lacks time reversal symmetry (TRS) or space inversion symmetry [1, 2]. For the TRS broken case, a Weyl semimetal (WSM) phase can be realized by a tunable parameter like magnetic order, and end up with a magnetic WSM [3–6]. It is then possible to engineer the magnetic order or domain structures to tune electronic and optical properties inherent in magnetic WSMs. Another route to WSM is to break the space inversion symmetry. In line with this strategy, the TaAs family was first theoretically predicted as the candidate materials and has later been experimentally confirmed to be a non-magnetic WSM [7–11]. However, the TaAs family as well as other predicted non-magnetic WSMs can hardly be tuned by experimentally accessible parameters to control the Weyl nodes or get rid of the trivial bands around the Fermi level, for realization of a clean band structure comprising Weyl nodes only around the Fermi level. These problems can be, however, overcome by exploiting the ferroelectricity as the inversion-symmetry breaking control parameter in the light of the topological phase diagram (see Fig. 1**a**) as proposed by Murakami and coworkers [12–15]. The coexistence of ferroelectricity and metallicity in a single material is usually unfavored due to the strong screening effect on the local electric dipoles. However, the concept of *ferroelectric metal* originally proposed by Anderson and Blount [16] has recently been substantiated by the discovery of many ferroelectric-like conductors, as exemplified by the ferroelectric metal $LiOsO_3$ [17] and electron-doped ferroelectric perovskite titanates [18, 19]. The semimetal state, regardless of being topological and non-topological, can be even a more suitable arena to endow with ferroelectricity, since the carrier density therein can be well tuned or minimized so as to be compatible with the ferroelectric order.

Figure 1**a** shows the schematic phase diagram based on Murakami's scheme specifically drawn for the case of the present target system $Pb_{1-x}Sn_xTe$, where the horizontal *x*- and vertical *y*-axes correspond to the gap tuning parameters such as Sn/Pb composition and the strength of inversion symmetry breaking like the ferroelectric order parameter, respectively. The phase diagram predicts that a WSM is unavoidably sandwiched between a normal insulator (NI) and a topological crystalline insulator (TCI) [20] when the inversion symmetry is broken by the ferroelectricity. The binary alloy $Pb_{1-x}Sn_xTe$ [21] serves as a key model system for realizing such topological phase transitions [22]. By composition variation between the two rocksalt-type compounds (space group No. 225; $Fm\bar{3}m$ when no ferroelectricity is induced), PbTe (NI) and SnTe

(TCI) [23], the mixed-crystal $Pb_{1-x}Sn_xTe$ shows a phase transition from NI to TCI at the critical composition $x_c \sim 0.35$ accompanied by a band inversion [24]. As for the symmetry-breaking lattice deformation, SnTe is known to be a ferroelectric [25] with a polar axis along <111> axes, while the ferroelectric transition tends to be suppressed by the increase in hole-type carriers. On the other hand, PbTe is an incipient ferroelectric, which shows a softened optical transverse phonon mode around 4 meV [26,27]. From these ferroelectric or incipient-ferroelectric behaviors in the both end compounds, it is likely that the mixed crystal $Pb_{1-x}Sn_xTe$ are also subject to ferroelectric instability (see the argument on the soft mode observed in the mixed crystal in Supplemental Information (SI) V). The bulk carrier concentration in $Pb_{1-x}Sn_xTe$, created possibly by atom deficiencies, can be well tuned by In doping in the relatively Pb-rich region [28], and eventually releases or revives the ferroelectric order in a specified composition range. Contrary to the case of the end SnTe, the polarity in the bulk crystal of $Pb_{1-x}Sn_xTe$ ($x\sim0.4$) is revealed to occur along the <001>, not <111>, directions by the present study via the second harmonic generation (SHG) anisotropy (*vide infra*).

The first-principles band calculations indeed predict the Weyl semimetal phase consistent with the Murakami's scheme for $Pb_{1-x}Sn_xTe$. The first Brillouin zones (BZ) without/with polar distortions are indicated in Fig. 1**b** and 1**c**, respectively. The W and K points on the hexagonal faces of the Brillouin zone (BZ, shown in Figs. 1**d** and 1**e**) become inequivalent, divided into $W_1/W_2$ and $K_1/K_2$, due to the loss of the $\mathcal{C}_3$ rotation and $\mathcal{M}_{yz}$ mirror symmetries when the [001] polar distortion appears. As shown in Fig. 1**i**, the band calculation shows that there appears an incipient phase (a critical phase between NI and WSM) when $x = 0.3$. Upon further tuning the Sn/Pb ratio or the polarity, each incipient touching point splits into one pair of Weyl nodes with opposite chirality (Fig. 1**e**), which are symmetric with respect to the vertical mirror plane ($\mathcal{M}_{xy}$). According to the Murakami's scheme [15], either WSM or nodal-line semimetal (NLSM) appears as the intermediate regime after the gap closing in inversion-symmetry-broken crystals, depending on the details of the irreducible representatives of the two bands involved in the topology. In particular, upon any phase transition between trivial insulator and topological insulator (including TCI), the intermediate phase should always have to be a WSM [13]. In the presence of inversion symmetry, at the critical Sn/Pb ratio, a band touching occurs at the center of the hexagonal facet of the BZ, *i.e.* the L point with $k_L = \frac{\pi}{2a} [1, 1, 1]$ (*a* being the lattice parameter), separating the NI from the TCI phase. Upon the development of the [001] polar distortion, all the bulk electronic states experience a potential gradient $\Delta V$ along this direction. This potential accordingly induces a spin-orbit field $\boldsymbol{B}_{so} \propto (\Delta V \times \boldsymbol{k})$, which at and near the L point is along the [-110] direction. Consequently, the initially spin-degenerate Fermi pockets become spin-polarized so that both the top valence band and bottom conduction band gain the same spin character, oriented along the [-110] direction. Since this field and resulting spin polarization are perpendicular to the (110) mirror plane, no nodal line encompassing the L point can be formed. Instead, the initial band touching point splits into two Weyl nodes, appearing along the $W_1 - L - W_1$ direction, parallel to [-110] direction, see Fig. 1**e**. When the bulk BZ without/with polar distortion projects onto the (001) surface, as indicated by red squares shown in Fig. 1**f** and 1**g**, the distribution of Weyl nodes on the surface BZ show $\mathcal{C}_4$ rotation symmetry and are symmetric with mirror $\mathcal{M}_{xy}$. Typical band dispersions (Figs. 1**h**-1**j**) of Pb$_{1-x}$Sn$_x$Te with the assumed [001] polar distortion show smooth band evolution from the NI phase to the TCI phase, while entering the WSM phase in-between. The projected surface states show the surface electronic structure for PbTe and SnTe are placed at the right part of Figs. 1**h**-1**j**, respectively, which confirms the distinct topologies; the helical surface state with Dirac cone characterized by mirror Chern numbers in the case of SnTe.

To confirm the scenario mentioned above and to detect the effect of Berry curvature in the WSM phase, we perform electrical and thermoelectric measurements for the series of indium doped Pb$_{1-x}$Sn$_x$Te samples; see Fig. 2**a** for a schematic diagram of the investigated samples [21]. Sample S36, located in the orange region with high mobility > $10^3$ cm$^2$/Vs shown in Fig. 2**a**, is selected as the representative WSM in the present study; sample S36 is shown to have the electric polarization along the [010] (*b*) axis as indicated via SHG measurements shown in Fig. 2**b** (experimental setup) and 2**c** (polar figure of SHG intensity). As shown in Fig. 2**d**, we place the current flow and thermal gradient along the *a*-axis, transverse Hall voltage contacts are prepared across b-axis, and magnetic field is applied along *c*-axis or *b*-axis whose geometry is referred to as the out-of-plane or in-plane Hall/Nernst configuration, respectively. As schematically shown in Fig. 2**e**, at zero magnetic field, Weyl nodes distribute in such a way that the total net Berry curvature is canceled out, as required by TRS. However, under applied magnetic fields, Weyl nodes re-distribute in the momentum space due to the Zeeman effect, allowing the effect of Berry curvature to be observed as field-induced anomalous Hall/Nernst signals, as described below. The first clue to Berry curvature generation is given by the observation of a sizable field-induced anomalous Hall effect $\sigma_{xy}^A$ with Hall angle as large as 0.04 in the in-plane field ($H \| b$) geometry as shown in Fig. 2**f**. In general, the total Hall conductivity can be expressed as $\sigma_{xy} = \sigma_{xy}^A + \sigma_{xy}^N$; $\sigma_{xy}^A$ is the anomalous component coming from the Berry curvature, and $\sigma_{xy}^N$ the conventional term from the Drude term. In the in-plane field configuration, importantly, no Lorentz force appears so that the conventional Drude term $\sigma_{xy}^N$ vanishes completely, allowing one to genuinely detect $\sigma_{xy}^A$ which originates from the Berry curvature. Here, if there were a mirror plane perpendicular to the *b*-axis, symmetry consideration requires that no in-plane Hall/Nernst signal should be observed, in other words $\sigma_{xy} = 0$ always holds [29]. However, as SHG measurements suggest that the polar domain is dominantly along the *b*-axis for S36 as shown in Fig. 2c, the mirror plane perpendicular to the *b*-axis is broken by the polarity. Therefore, the finite in-plane Hall signals, i.e., $\sigma_{xy}^A$, are allowed to emerge for S36. For the conventional out-of-plane geometry, on the other hand, the Drude term $\sigma_{xy}^N$ also shows up and the Hall conductivity to be detected is $\sigma_{xy} = \sigma_{xy}^A + \sigma_{xy}^N$ with $\sigma_{xy}^N = pe\mu^2B/(1+\mu^2B^2)$, where *p* is the carrier concentration, *e* the elemental charge and μ the carrier mobility. The fitted result shows the $\sigma_{xy}^A$ component indicated by the pink shadowed area in Fig. 2**f**, whose magnitude is comparable with that of $\sigma_{xy}^A$ detected in the in-plane geometry as expected.

While $\sigma_{xy}^A$ can in principle be extracted via such an elaborate subtraction procedure in the out-of-plane field geometry, the large background of $\sigma_{xy}^N$ hinders the accurate experimental detection of the anomalous term originating from the Berry curvature. In order to directly detect the effect of the Berry curvature for the out-of-plane geometry, we consider the Nernst effect as the more sensitive thermoelectric probe; Nernst effect measures the ratio (Nernst coefficient $S_{xy}$) of the transverse (//*y*) voltage drop $V_y$ under longitudinal thermal current or temperature gradient (//*x*) $\nabla_x T$ under external magnetic field, as shown in Fig. 2**b**. With the use of the Seebeck coefficient $S_{xx}$ as well as $S_{xy}$, the transverse thermoelectric tensor $\alpha_{xy}$ related to the Nernst effect can be written as $\alpha_{xy} = [\sigma_{xx}S_{xy} + \sigma_{xy}S_{xx}] = \frac{S_{xy}}{\rho_{xx}} + \frac{\rho_{yx}}{\rho_{xx}}\alpha_{xx}$, where $\alpha_{xx} = [\sigma_{yx}S_{xy} + \sigma_{xx}S_{xx}]$ and $\rho_{ij}$ is the electric conductivity tensor satisfying the relation, *e.g.*, $\sigma_{xy} = \frac{\rho_{yx}}{\rho_{xx}^2 + \rho_{yx}^2}$. Again decomposing $\alpha_{xy}$ into the anomalous and normal component, $\alpha_{xy} = \alpha^A_{xy} + \alpha^N_{xy} \sim \alpha^A_{xy}$, where the contribution from Drude term $\alpha^N_{xy}$ is known to be greatly suppressed given the Mott relation, $\alpha_{xy}(= -\frac{\pi^2 k_B^2}{3|e|}\frac{\partial \sigma_{xy}}{\partial \mu})$ with small $\frac{\partial \sigma^N_{xy}}{\partial \mu}$ term[30-33]. Hence, by focusing on $\alpha_{xy}$, the effect of Berry curvature can be commonly extracted both for the out-of-plane and in-plane configurations. Indeed, as shown in Fig. 2**g**, the magnitude of $\alpha_{xy}$ is comparable for both the out-of-plane and in-plane geometries unlike the case for $\sigma_{xy}$.

To gain further insight on the Berry curvature, we analyze the details of the in-plane field effects. As already mentioned, the in-plane field geometry can extract the genuine contribution of Berry curvature from both $\sigma_{xy}$ and $\alpha_{xy}$ results. Figures 3**a** and 3**b** show the field dependence of the in-plane Hall conductivity ($\sigma_{xy}^{\text{in-plane}}$) and the in-plane transverse thermoelectric conductivity ($\alpha_{xy}^{\text{in-plane}}$) at selected temperatures; the 300 K data for $\sigma_{xy}^{\text{in-plane}}$ serve as a null reference, while the $\alpha_{xy}^{\text{in-plane}}$ signal vanishes at temperatures above 100 K. For $\sigma_{xy}^{\text{in-plane}}$, the expression reads [22],

$$\sigma_{xy}^{\text{in-plane}} = \sigma^A_{xy} = \frac{e^2}{\hbar} \int \frac{d\mathbf{k}}{(2\pi)^3} \Omega_z(\mathbf{k}) f^0_k = e\langle\Omega_z\rangle \cdot p \quad, \tag{1}$$

where $f^0_k$ is the Fermi-Dirac distribution, $\langle\Omega_z\rangle$ the averaged Berry curvature over the BZ, and $p$ the whole carrier density. As shown in Eq. (1), the averaged Berry curvature is $\langle\Omega_z\rangle \sim \sigma_{xy}^{\text{in-plane}}/p$ which is plotted in blue in Fig. 3**d**; for the plot of temperature dependence, the thermal broadening of $f^0_k$ is ignored here. For $\alpha^A_{xy}$ holds the Mott relation, $\alpha^A_{xy}(T,\mu) = -\frac{\pi^2 k_B^2}{3|e|}\frac{\partial \sigma^A_{xy}}{\partial \mu}$, where $k_B$ is the Boltzmann constant, $e$ the elemental charge and μ the mobility. Then we obtain

$$\alpha_{xy}^{\text{in-plane}}(T,\mu) = \alpha^A_{xy}(T,\mu) = \frac{e^2}{\hbar}\int \frac{d\mathbf{k}}{(2\pi)^3}\Omega_z(\mathbf{k})w(E - E_F), \tag{2}$$

where, $w(\epsilon) = -\frac{1}{eT}[\epsilon f^0_k + k_B T \ln(1 + \exp(-\epsilon/k_B T))]$ is the weighting factor, which picks up both occupied and unoccupied states but only around Fermi level. Via analyses similar to Eq. (1) [30–32], we find

$$\frac{\alpha_{xy}^A}{T} \sim -\frac{m^* k_B^2}{\hbar^2} \langle \Omega_z \rangle \cdot p^{\frac{1}{3}} \quad , \tag{3}$$

where $m^*$ is the effective mass. From Eq. (3), the averaged Berry curvature $\langle \Omega_z \rangle \propto \frac{\alpha_{xy}^A}{T}/p^{1/3}$, which is also plotted in red in Fig. 3d. As shown in Fig. 3d, the $\langle \Omega_z \rangle$ values extracted from the in-plane signals, $\sigma_{xy}^{\text{in-plane}}$ and $\alpha_{xy}^{\text{in-plane}}$, show nearly identical temperature dependence. The signals start to increase below around 100 K where the magnitude of inversion symmetry breaking as evaluated by SHG intensity (Fig. 3**c**) appears to develop enough to bring the system into the WSM phase. From Eqs. (1) and (3), the ratio $\alpha_{xy}^A/T\sigma_{xy}^A$ in the in-plane configuration is equal to $(m^* k_B^2/e\hbar^2) * p^{-2/3}$. As shown in Fig. 3e, the experimental values of the ratio $\alpha_{xy}^A/T\sigma_{xy}^A$ calculated from the two in-plane field quantities are comparable with the theoretical ratio if we assume the effective mass as $m^* = 0.1\ m_e$. The consistency between experiment and theory shows the correctness of the picture based on Berry curvature.

To reveal the response of Weyl nodes with respect to the polar distortion and applied magnetic fields, we performed full field-rotation Nernst measurements on more samples (S33, S35, S36, and S40). Samples S33 and S40 are typical samples located in the Sn-rich side with little SHG signal and low mobility as indicated by the blue shadow region in Fig. 2**a**. By contrast, samples S35 and S36 are located close to Pb rich side and show large SHG signals with moderately high mobility (for details of material characterizations, see Ref. [21]). Here, we focus on the comparison of S36 (WSM) and S33 (non-WSM); samples S35 (WSM) and S40 (non-WSM) respectively show similar Nernst features correlated with the SHG activity to S36 and S33 (for details, refer to the SI). Insets to Figs. 4**a**-**c** show schematic illustrations of measurement configurations and definition of angles. As shown in Eq.3, in the strict sense, the effect of Berry curvature should be evaluated by $\alpha_{xy}/T$, rather than the Nernst coefficient $S_{xy}$. However, because of a practical technical difficulty in obtaining $\alpha_{xy}$ via measurements during the continuous sample rotation (see SI for details), the Nernst signals divided by resistivity at zero magnetic field ($\rho_0$, *i.e.*, $\rho_{xx}$ at $H=0$) and temperature (*T*), *i.e.*, $S_{xy}/\rho_0 T \sim \alpha_{xy}/T$ (see SI section I for details) is used as the alternative to evaluate the effect of Berry curvature. The comparison of S33 and S36 explicitly shows that the origin of the Weyl nodes is the polar distortion, and not the applied magnetic fields. In other words, the Weyl nodes already exist at zero magnetic field. This is evidenced by the fact that only the ferroelectric samples like sample S36 (and S35) show the sizable effect of Berry curvature. Furthermore, without polar distortion, from the viewpoint of mirror symmetry, no in-plane signals can appear when H//b. In reality, sizable in-plane signals even when H//b are observed for sample S36, clearly evidencing that Weyl nodes already exist at zero magnetic field due to the polar distortion. Having revealed the ferroelectric order as the origin of the Weyl nodes, we then briefly discuss the details of the angular dependence of the Berry curvature generation under applied magnetic fields, as shown in Figs. 4**e**-**g**. Here, the key ingredient is the two mirror planes perpendicular to the *a* axis and *b* axis. In order to generate finite $\alpha_{xy}^A$ (or Berry curvature), both mirror planes must be broken. As already mentioned, polarization is predominantly along *b* axis according to the SHG measurement shown in Fig. 4**h** for this sample, breaking one of the relevant mirror planes. The other mirror plane perpendicular to *a* axis is also broken whenever *H* is directed away from *a* axis, generating finite amount of Berry curvature, as exactly shown in Figs. 4**e**-**g**. Here, we note that the finite small Berry curvature even when H//a is due to the fact that in reality S36 also contains a small fraction of polar domains with polarization along *a* axis.

In conclusion, we find a non-magnetic WSM in the In-doped $Pb_{1-x}Sn_x$Te system coupled with the intrinsic ferroelectric order, which is tunable by adjusting In doping concentration and Sn/Pb ratio. The ferroelectric WSM phase shows a clean band structure around Fermi level, providing an ideal platform for investigating properties of Weyl fermions. A newly established transport probe, namely the field-angle dependent Nernst effect, precludes the contribution from the conventional Drude term both for the out-of-plane and in-plane geometries, and can be used to selectively detect the effect of Berry curvature which is generated by the re-distribution of Weyl nodes under applied magnetic fields. This provides a new step towards merging the concepts of ferroelectricity and band topology. The highly tunable ferroelectric order will bring a plethora of Weyl fermion physics, such as turning on/off the Weyl phase, varying the separation length of Weyl nodes in momentum space, and adjusting the topological band structures around Fermi level.

## Methods

### Crystal growth and characterizations

Single crystals of $(Pb_{1-x}Sn_x)_{1-y}In_y Te$ with various $x$ and $y$ values were grown by the conventional vertical Bridgman-Stockbarger technique. We used high-purity elements (Pb, Sn, In, Te; 5N) with a prescribed stoichiometric ratio and sealed them into carbon pre-coated fused silica ampoules. The ingredients were pre-melted and rocked in a box furnace, and then transferred to the Bridgman furnace. The temperatures for the upper and lower furnaces were 980 °C and 650 °C, respectively. The ampoules were slowly pulled down at a speed of 1 mm/hour to obtain the crystal rod with some gradient of the composition along the growth direction [21]. The crystal rods were cut into thin (~1 mm) plates. Powder x-ray diffraction (XRD, Rigaku), Laue (Rigaku) measurements and energy dispersive x-ray (EDX) were subsequently performed for each plate to confirm the crystal structure, crystallographic directions, and actual element composition, as described in detail elsewhere [21].

### Transport measurements

Electrical transport measurements were carried out in a Quantum Design Physical Property Measurement System (PPMS) equipped with a sample rotator. A standard four-probe method was employed by using silver paste and ultrasound indium welding for making electrical contacts. All the samples were cut into thin rectangle shapes, and then polished for transport measurements.

Thermoelectric measurements of out-of-plane field configurations are performed on a home-built setup, namely one-heater and two-thermometer scheme, as shown in Fig. S1(a). Here, we use thermocouples to monitor the temperature gradient. The heater is connected by Keithley AC/DC current source 6221 and the longitudinal (Seebeck) and transverse (Nernst) signals are simultaneously monitored by two Keithley 2182a nanometers. All the thermal measurements are carried out with PPMS in a vacuum environment. For the angular-dependent thermoelectric measurements, we slightly modified the PPMS rotator (for details see Fig. S1 (a)) to manually build up a good heat sink, then we constructed the same thermoelectric setup as mentioned above. The definition of the sign of the Nernst effect is chosen so as to be consistent with the custom described in Refs. [34,35].

**Optical second harmonic generation measurements**

Optical second harmonic generation (SHG) measurements were performed by using a pulsed laser source (120 fs, 1.55 eV). We adopted a configuration of normal incidence on the *ab* plane to detect in-plane polar distortion by ruling out the parasitic surface-normal contribution. The polarization of the laser was rotated by a half-wave plate, and reflected SHG signal was analyzed by a Glan-laser prism. The temperature dependent experiment was carried out in a high-vacuum cryostat down to 10 K.

**First-principles calculation**

To study the electronic structure of $Pb_{1-x}Sn_xTe$, we first performed density functional theory (DFT) calculations for the end compounds SnTe and PbTe, individually, using the Perdew-Burke-Ernzerhof (PBE) exchange-correlation functional [36] and ultrasoft pseudopotentials, implemented in the VASP programme [37]. To incorporate the effect of [001] polar distortion, all the Te atoms were rigidly shifted (by 0.1% in the case of SnTe and 3% in the case of PbTe), along this direction, while keeping the lattice parameters fixed to their experimentally reported values, $a = 6.31$ Å for cubic SnTe and $a = 6.46$ Å for cubic PbTe [38]. The corresponding Brillouin zone was sampled by a $20 \times 20 \times 20$ k-mesh. From the DFT Hamiltonian, we constructed a 12-band tight-binding (TB) model using maximally localized Wannier functions [39], considering Pb/Sn-$5/6p$ and Te-5p as the projection centres. We then used a linear interpolation of these two TB Hamiltonians to model the electronic structure of the solid solution $Pb_{1-x}Sn_xTe$.

**Acknowledgments** We thank M. Hirayama, M. Kawasaki and N. Nagaosa for helpful discussions. This project was partly supported by CREST (Grant number, JPMJCR16F1) from JST. N.O. is supported by PRESTO JST (No. JPMJPR17I3).

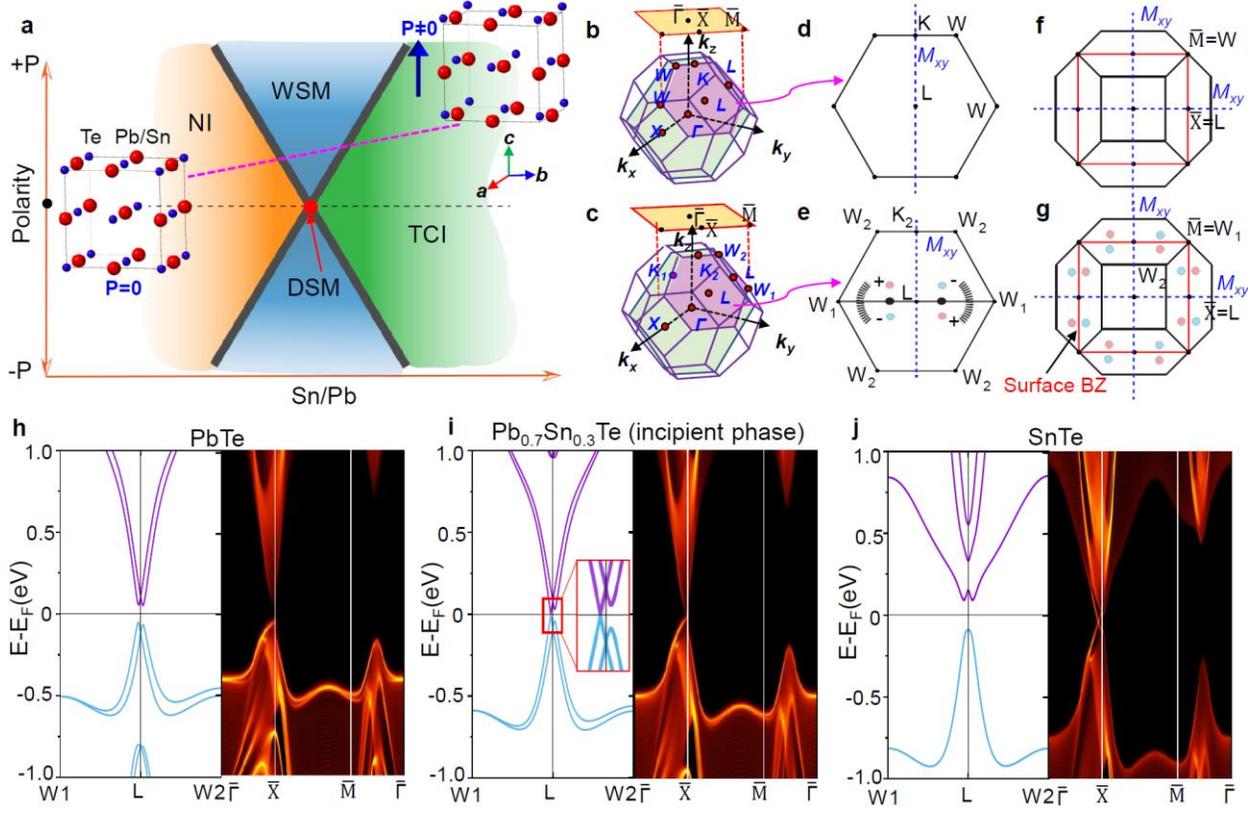

Fig. 1| **A ferroelectric Weyl semimetal arising from PbTe-SnTe alloy as revealed by first-principles calculations. a,** Topological phase diagram based on Murakami's scheme, in which topologically-distinct phases in $Pb_{1-x}Sn_xTe$ alloy are controlled by two parameters, namely polarity and Sn/Pb ratio. Insets show crystal structures of $Pb_{1-x}Sn_xTe$ with/without [001] polar distortion, respectively. **b&c,** First Brillouin zone without/with [001] polar distortion, respectively. Yellow shadowed areas are the projected (001) surface Brillouin zones. **d&e,** A top view of one of the hexagonal faces, shaded by violet in **b** and **c**. Blue dashed line shows the mirror plane $\mathcal{M}_{xy}$, and the emergent Weyl nodes are located symmetrically alongside $\mathcal{M}_{xy}$ with opposite chirality. **f&g,** Weyl nodes on the projected (001) surface Brillouin zone (denoted as red squares) without/with [001] polar distortion, respectively. **h-j,** Bulk (left) and surface (right) band structures for the two end compounds (PbTe and SnTe) as well as the incipient Weyl semimetal phase ($Pb_{0.7}Sn_{0.3}Te$) with the hypothetical polarization along [001]. Note here that the polar axis is set along [001], as one of the possible equivalent <100> axes. Hereafter, the dominant polar axis in the sample (S36) investigated in this study is defined as the **b** axis.

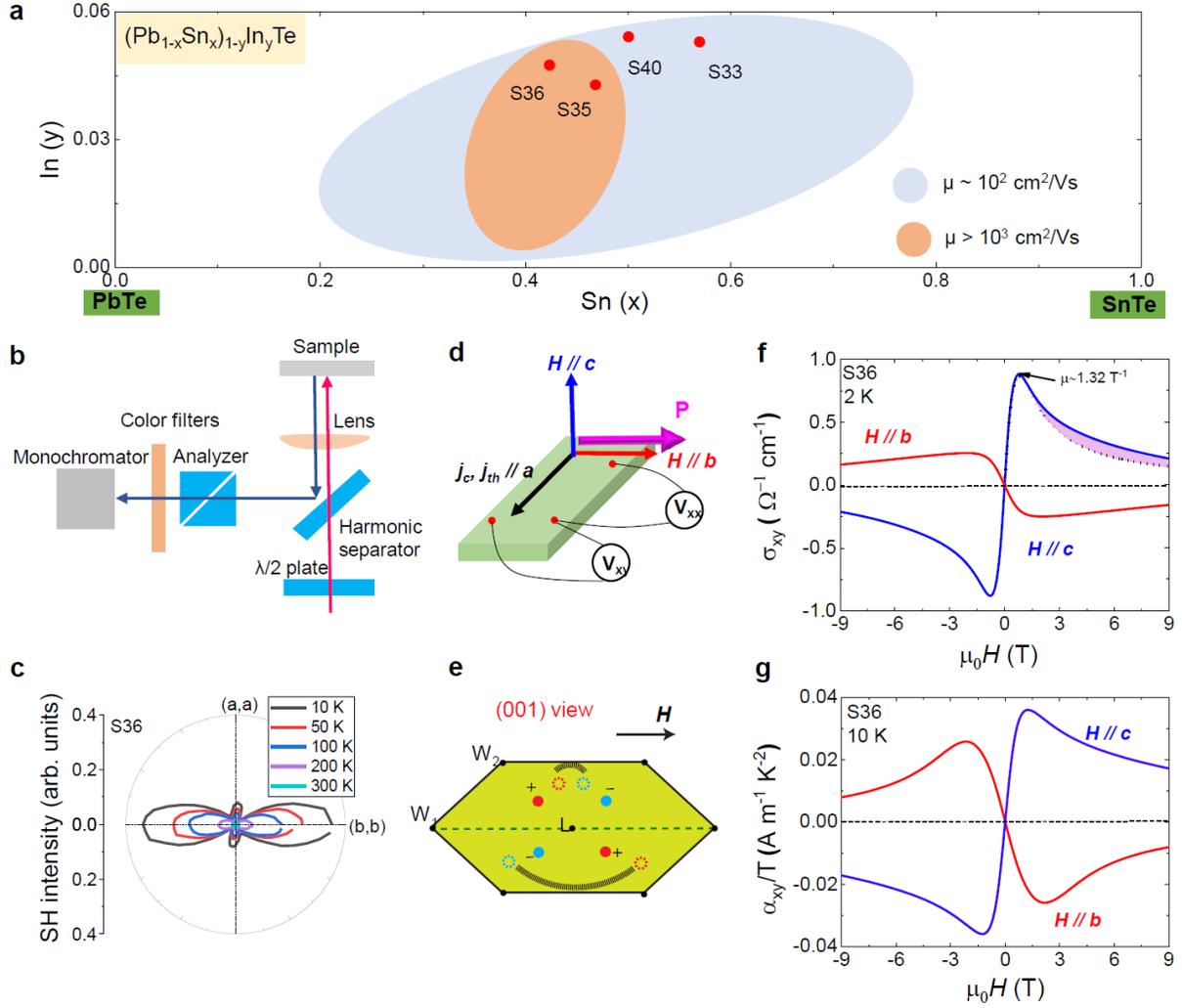

Fig. 2| **Field-directional anisotropy of $\sigma_{xy}$, $\alpha_{xy}/T$ in the Weyl semimetal phase. a,** The schematic carrier mobility contour map from the previous study [21], and the locations of the targeted samples in this paper. **b,** An illustration of second-harmonic-generation (SHG) measurement setup. The fundamental light is incident on the (001) plane. **c,** Light-polarization and temperature dependent SHG signals plotted as a polar figure for sample S36. For example, (b, b) stands for the fundamental and second-harmonic light polarizations in order. In this definition of the axes, the polarity (**P**) mainly directs along the b axis. **d,** Configuration of the electrical/thermoelectric measurements. The in-plane configuration denoted as **H**||**b** and out-of-plane configuration denoted as **H**||**c**. **e,** A schematic hexagonal face in (001) view where the Weyl nodes are re-distributed when magnetic field along the b axis (polar axis) is applied. **f,** The out-of-plane (**H**||**c**)/in-plane (**H**||**b**) Hall conductivity versus magnetic field at 2 K. The black arrow

indicates the resonant peak, representing the high mobility around ∼ 1.32 T$^{-1}$ ($1.32 \times 10^4$ cm$^2$/Vs). The single-band (Drude) model fits well to the 2 K out-of-plane data in low fields, while it exhibits deviation in higher fields, as indicated by the area shaded in pink. The excess conductivity denotes the anomalous term arising from the emergent Berry curvature under the magnetic field. **g,** $\alpha_{xy}/T$ measured in out-of-plane (**H**||**c**) and in-plane (**H**||**b**) geometry for sample S36 at 10 K. The observed signals for the two configurations show comparable values, both of which come from the Berry curvature.

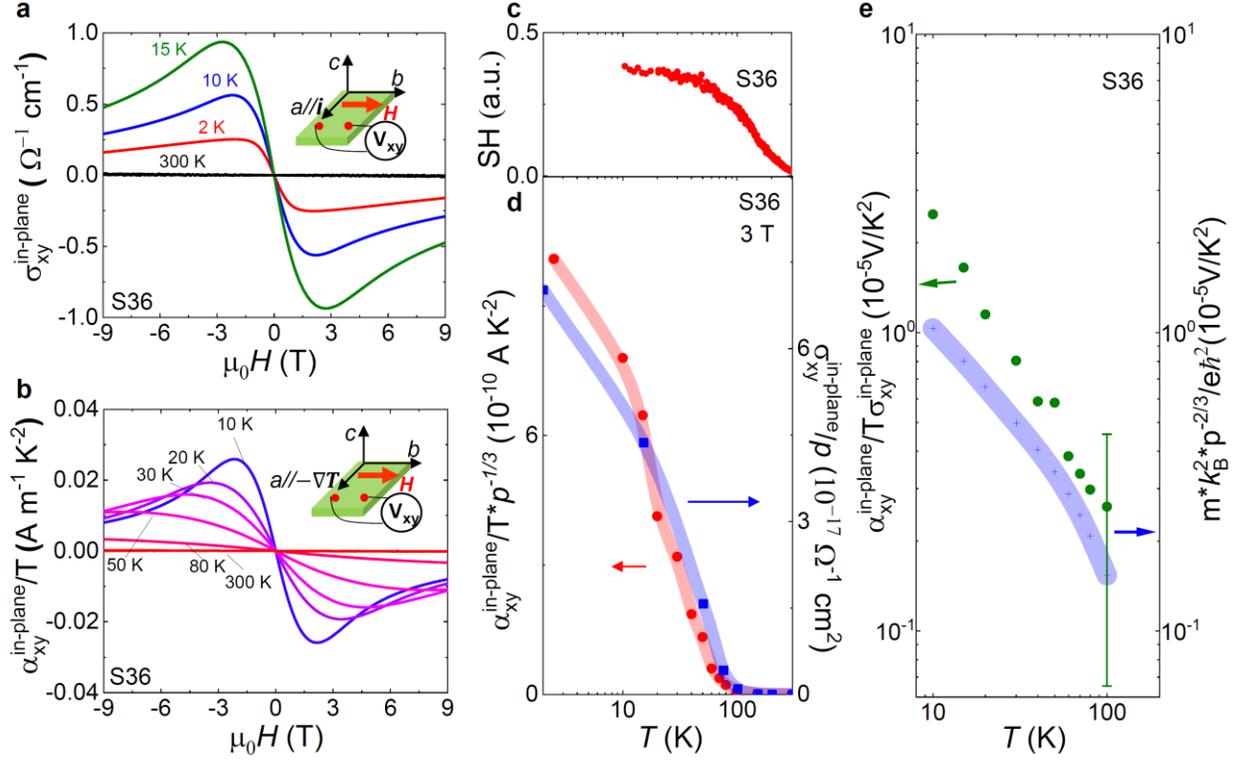

**Fig. 3| In-plane Hall ($\sigma_{xy}^{\text{in-plane}}$) and Nernst ($\alpha_{xy}^{\text{in-plane}}/T$) responses arising from Berry curvature.** **a,** In-plane Hall conductivity versus magnetic field at selected temperatures. Inset shows the experimental configuration for the Hall measurement. **b,** In-plane $\alpha_{xy}/T$ versus magnetic field at selected temperatures. Signals almost vanish at temperatures above 100 K. Inset shows the experimental configuration for the $\alpha_{xy}/T$ measurement. **c,** Temperature dependence of optical second-harmonic (SH) signal in the SHG experiment, showing that the electric polarity grows as the temperature decreases. **d,** Two rescaled quantities, $\sigma_{xy}^{\text{in-plane}}/p$ and $\alpha_{xy}^{\text{in-plane}}/T\,p^{1/3}$ proportional to the averaged Berry curvature $<\Omega_z>$, show large enhancement at temperatures below ~ 100 K, where the magnitude of inversion symmetry breaking (polarity) becomes large enough to bring the system into WSM phase. $p$ represents the carrier density. **e,** The ratio of the in-plane field quantities, $\alpha_{xy}/T\sigma_{xy}$, versus the predicted theoretical value for the Berry curvature contribution. Colored thick lines in **d** and **e** are the guides to the eyes.

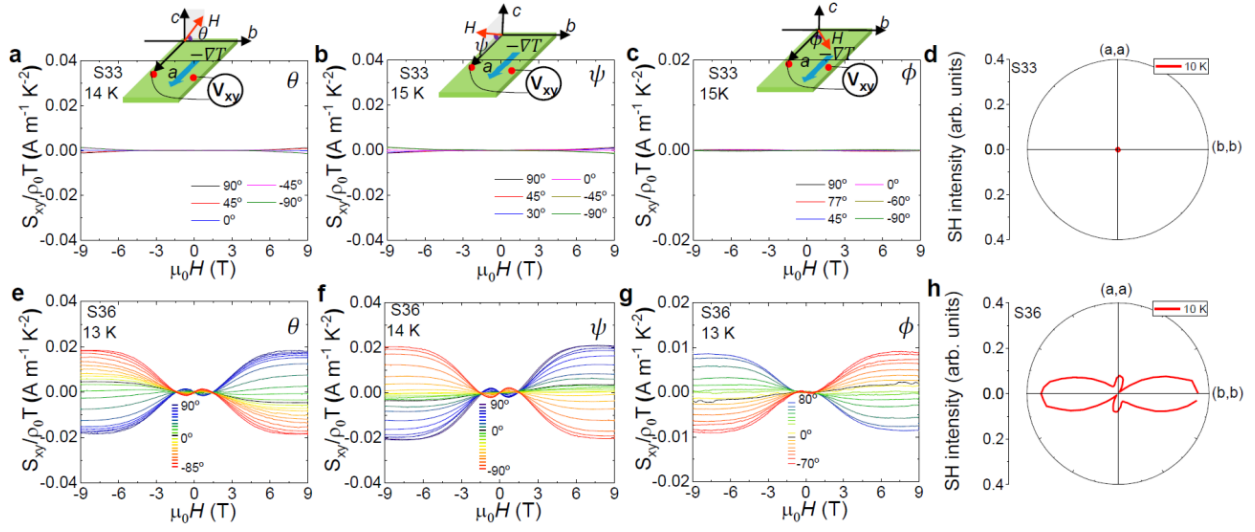

Fig. 4| **Field angular dependent Nernst effect and relation with spontaneous polar distortion.**

Field-rotational variation of Nernst effect $S_{xy}/\rho_0 T$ (scaled by longitudinal resistivity $\rho_0$) in sample S33 (**a-c**) and sample S36 (**e-g**), respectively. The sizable Nernst effect $S_{xy}/\rho_0 T$, which reflects the effect of Berry curvature generated by the re-distribution of Weyl nodes under applied magnetic fields, is observed only for sample S36. Insets show three configurations of applied temperature gradient and magnetic field with the field angles (θ, ψ, φ). **d&h** Corresponding SHG polar figure plots at 10K, plotted on the same relative-magnitude scale.

20